\documentclass{mecbic}
\providecommand{\anno}{2011}
\providecommand{\firstpage}{113}
\providecommand{\lastpage}{114}
\usepackage[english]{babel}
\usepackage{makeidx}  
\usepackage{graphicx}
\usepackage{amsfonts}
\usepackage{wrapfig}
\usepackage{enumitem}
\usepackage[T1]{fontenc}
\usepackage{subfigure}

\title{Multiscale Modelling: A Mobile Membrane Approach \\  {\large (Work in Progress)}}
\author{Federico Buti, Massimo Callisto De Donato, Flavio Corradini, Emanuela Merelli, Luca Tesei
\institute{School of Science and Technology, Computer Science Division\\
University of Camerino, Camerino, Italy
\\}
\email{\{name.surname\}@unicam.it}
}
\def\titlerunning{Multiscale Modelling: a Mobile Membrane Approach}
\def\authorrunning{F. Buti, M. Callisto De Donato,\\\ F. Corradini, E. Merelli and L. Tesei}

\begin{document}
\maketitle      
\pagestyle{plain}
\pagenumbering{arabic}
\setcounter{page}{113}  
\begin{abstract}
Nowadays, multiscale modelling is recognized as the most suitable way to study biological processes. Indeed, almost every phenomenon in nature exhibits a multiscale behaviour, i.e., it is the outcome of interactions that occur at different spatial and temporal scales. Although several ways to provide ``multilayer'' models have been proposed, only Complex Automata naturally embed spatial information and realize the multiscale approach with well-established inter-scale integration schemas. Recently, such approach has been restated in terms of Spatial P systems - a variant of P systems with a more geometric concept of space.

\noindent In this work we discuss how mobile membranes, a variant of membrane systems inspired by the biological movements of endocytosis and exocytosis, can be efficaciously exploited to define a uniform multiscale coupling scheme relying only on the features of the formalism itself. 

\end{abstract}
\section{Introduction} \label{sec:intro}
Many biological phenomena are characterized by interactions involving different spatial and temporal scales simultaneously. At each scale, different structures come into play. Consequently, several computational methodologies have been developed for modelling biological processes with different degrees of resolution. Among them, {\it multiscale} modelling rose up as the most suitable one. In such approach, several models at different spatial and temporal scales, are \textit{coupled} together. How these models are homogenized, i.e, \textit{integrated} is a key aspect to guarantee that the overall model is faithful to the real phenomenon.

Recently, complex Automata (CxA) \cite{HoekstraFCC08} were introduced as a robust multiscale modelling solution which takes in account coupling issues. In CxA the multiscaling is realised on uniform components, the {\em Cellular Automata}, (CA) 
by different and well-established integration schemas. Any CxA consists of a finite grid of CA cells, where each cell has an associated state taken from a finite set of different states. 
In \cite{CCMT10}, authors rephrased the CxA approach in term of Spatial P Systems (SPs)~\cite{sps}, a variant of P Systems enriched with a notion of space. Similarly to CxA, SPs represents space as a geometric 2D grid-based set of cells that contain objects as in classical P Systems.
As a case study, authors provided a uniform multiscale model of bone remodelling, a common multiscale phenomena, as an aggregation of single SP models at different scales, coupled through functions \textit{external} to the paradigm. 
Following their approach, we discuss how mobile membranes\cite{Ciobanu2010} can be exploited to define a multiscale coupling scheme which relies only on the operations provided from the mobile membrane paradigm itself.

\section{Bone remodelling with mobile membranes} \label{sec:bone}
\begin{figure}
	\begin{center}
		\subfigure[MODEL]{
		  \includegraphics[width=0.36\textwidth]{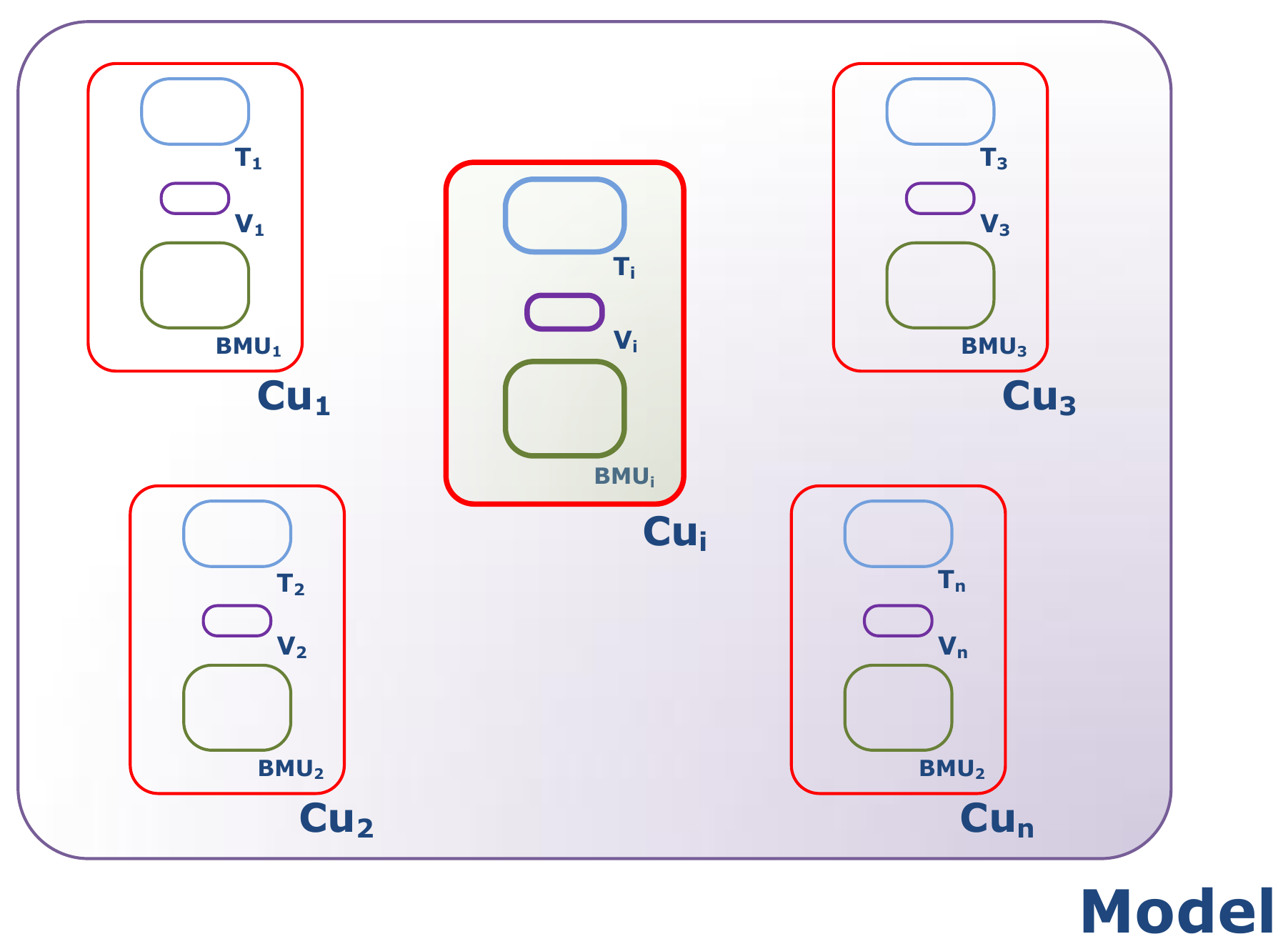}
		  \label{fig:model}
		}\qquad
		\subfigure[BMU]{
		  \includegraphics[width=0.34\textwidth]{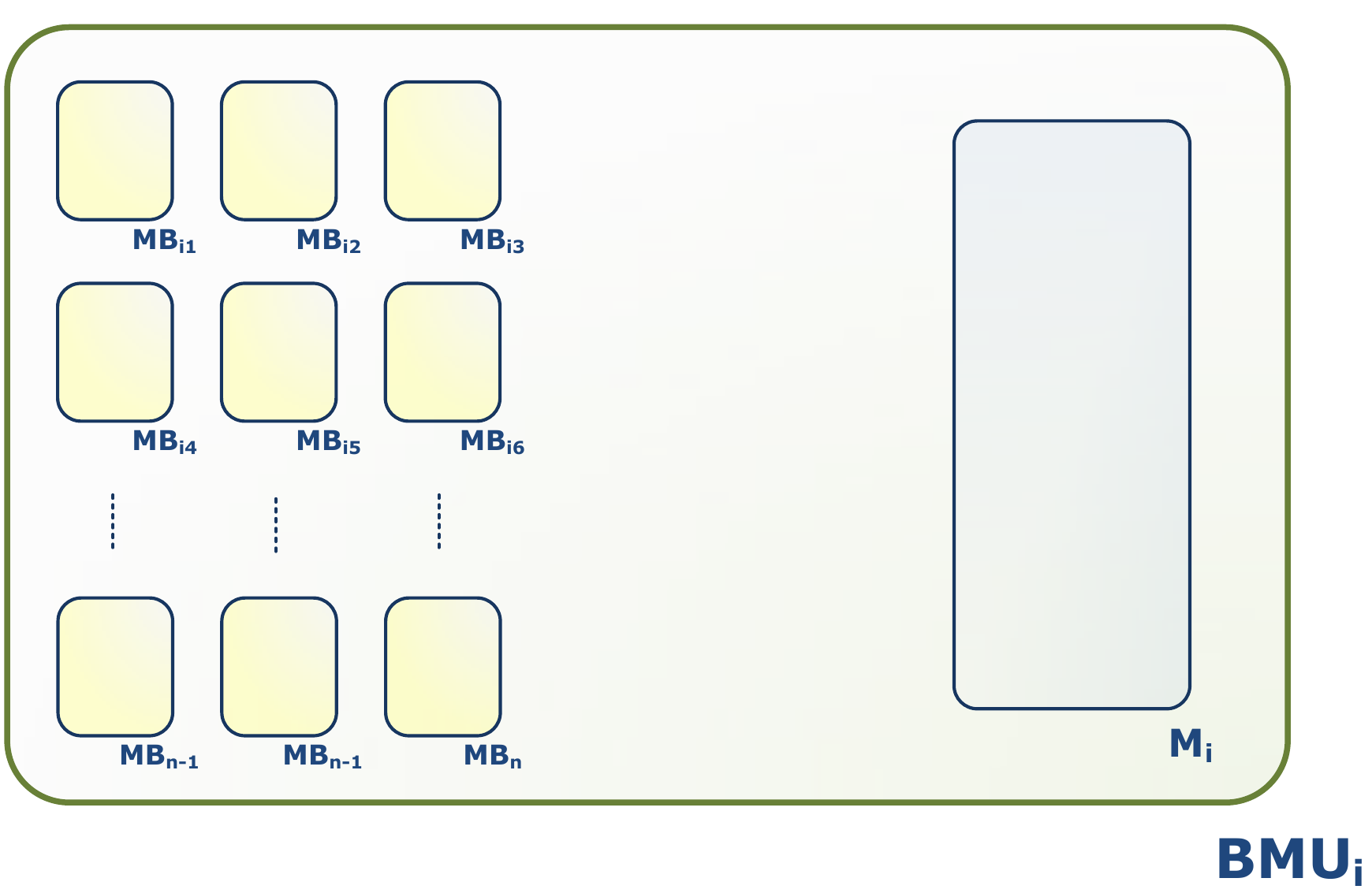}
		  \label{fig:bmu}
		}
\end{center}
\caption{Multiscale model for bone remodelling. Figure (a) represents the whole model containing the coupling membranes CU (red).  Figure (b) shows a detailed BMU (cell scale) membrane.}
\label{fig:bonemmm}
\end{figure}

Bone remodelling is a biological phenomenon in which mature bone tissue is removed from the skeleton (\textit{resorption}) and new bone tissue is formed (\textit{formation}); such process is multiscale since macroscopic behaviours, e.g. mechanical stimuli at the tissutal level and microstructure (cell scale actors - e.g. hormones, receptors etc.) strongly influence each other.
Figure~\ref{fig:model} shows the proposed model based on mobile membranes. As in \cite{CCMT10}, we consider two resolution levels: the tissutal and the cellular ones. Tissue level is divided into a series of membranes $T_i$ (cyan), each representing a portion of bone. Each membrane contains, at any moment, a number of objects $c$ proportional to the mineralisation value (expressed as a density in a certain interval) of the tissue. At the cellular scale we consider a series of $BMU_i$ membranes (green), i.e. ``Bone Multicellular unit'' membranes representing the cellular sites in which remodelling occurs\footnote{A detailed description can be found at: \url{http://courses.washington.edu/bonephys/physremod.html}.}. The integration of the two scales is realised through a coupling membrane $CU_i$ (red) and a simple membrane $V_i$ (violet), that moves between the tissutal and cellular membranes carrying the coupling information. No external function is needed since the paradigm provide the right means to handle the coupling.

\vspace{-2ex}

\section{Future work}\label{conclusion}
We sketched a possible application of Mobile Membranes to the definition of a multiscale coupling scheme which, differently from \cite{CCMT10}, totally relies on the paradigm itself. In the near future we aim to fully develop this approach. Moreover,  we also aim to study the paradigm expressivity w.r.t. CxA, and SPs. To this aim, an extension of Mobile Membranes with an explicit spacial notion is mandatory. 

\bibliographystyle{eptcs}
\bibliography{MMM}
\end{document}